# Permanent and Transient Synchronized Chaos in Large Arrays of Complex-Coupled Semiconductor Lasers

ZHANNING LIU, HERBERT G WINFUL*

*Department of Electrical Engineering and Computer Science, University of Michigan, 1301 Beal Avenue, Ann Arbor, Michigan 48109-2122*

*arrays@umich.edu

**Abstract:** Synchronized chaos has previously been predicted and observed in a small number (3) of mutually coupled lasers. In this work, we demonstrate that this phenomenon can theoretically persist in significantly broader scenarios, extending to complex coupled arrays of up to 11 lasers and arrays with finite built-in disorder. We quantify the resulting high-dimensional dynamics by computing Lyapunov spectra and the associated Lyapunov dimension, confirming that the observed states are chaotic rather than quasi-periodic. Furthermore, we uncover a regime of transient synchronized chaos where the system eventually escapes from perfectly synchronized chaotic state into an asynchronous state. We find that the lifetime of these transient states follows a bi-exponential distribution.

## 1. Introduction

Coupled semiconductor laser arrays provide a rich framework for studying high-dimensional nonlinear dynamics, where chaos emerges from intricate mode interactions [1]. In particular, synchronized chaos arises when individual lasers in the array exhibit chaotic temporal behavior yet remain perfectly synchronized in phase and amplitude [2, 3]. Possible applications of such synchronized chaos include secure communications, neural networks, the generation of random bits, secure key distribution, and machine learning [4]. While this phenomenon has been studied theoretically [2, 3], [5-8] and experimentally [9] for 3 – 5 coupled semiconductor and solid-state laser arrays, it is not clear how it scales to large numbers. This is because the number of dimensions in phase space available for the system to explore grows rapidly, typically as 3N, where N is the number of lasers in the array. The effect of complex coupling, detuning, and unequal pumping further expands this landscape.

In this paper, we model the array using the Winful-Wang coupled-mode equations [1] and demonstrate the existence and robustness of synchronized chaos over microsecond timescales in arrays of up to 11 identical complex-coupled lasers, as well as in 3-element arrays with built-in parameter disorder. Furthermore, by exploring the regime of large coupling phases, we uncover a novel dynamical state: transient synchronized chaos, in which the system perfectly synchronizes before ultimately breaking its spatial symmetry and evolving into asynchronous dynamics. While transient chaos is a well-studied phenomenon [10,11], transient synchronized chaos has received little attention [3]. We characterize this behavior and find that the lifetime of the transient states is governed by a bi-exponential distribution.

## 2. Theoretical Formulation

Our model is a 1-dimensional open chain array of waveguide lasers coupled evanescently. The dynamics of the jth laser in the array of N coupled semiconductor lasers are described by the Winful-Wang coupled mode equations [1] where the slowly varying field amplitude, phase, and normalized carrier density are $X_j$, $\phi_j$, and $Z_j$ respectively:

$$\dot{X}_j = Z_j X_j - |\eta|\left[X_{j+1}\sin(\theta_j + \psi) - X_{j-1}\sin(\theta_{j-1} - \psi)\right]$$

$$\dot{\phi}_j = \Delta_j + \alpha Z_j - |\eta|\left[\frac{X_{j+1}}{X_j}\cos(\theta_j + \psi) + \frac{X_{j-1}}{X_j}\cos(\theta_{j-1} - \psi)\right]$$

$$T\dot{Z}_j = p_j - Z_j - (1 + 2Z_j)X_j^2$$

Here $\eta = |\eta|e^{i\psi}$ is the complex coupling parameter, $\alpha$ is the linewidth enhancement factor, $p_j$ is the normalized excess pumping rate, $\Delta_j$ is the detuning of the jth laser compared to a common reference optical frequency, $t$ is the time normalized by photon lifetime ($1.6 * 10^{-12}$ s is assumed throughout this paper), and absorbing boundary condition $X_0 = X_{N+1} = 0$ is applied.

For a fixed set of parameters, if we increase the coupling strength between neighboring laser pairs, the system gives rise to a series of bifurcations, from Hopf bifurcation to a region of full chaos [12]. Such a bifurcation to chaos is shown in Figure 1 for 3 coupled lasers with complex coupling. In this figure, the red triangles represent the extrema of the field amplitude in laser for an asymmetric initial phase distribution while the black dots are for a symmetric initial condition. It was found that for an odd number of lasers, there exists a symmetry-preserving state where symmetrically located pairs are identical in their temporal evolution. Synchronized chaos stems from the fact that the stability of this symmetry-preserving state persists for a certain distance along the coupling strength axis even after the temporal behavior of the laser becomes fully chaotic. The range of coupling that induces synchronized chaos can be understood as regions between the turning on of chaos and symmetry breaking in the spatial domain.

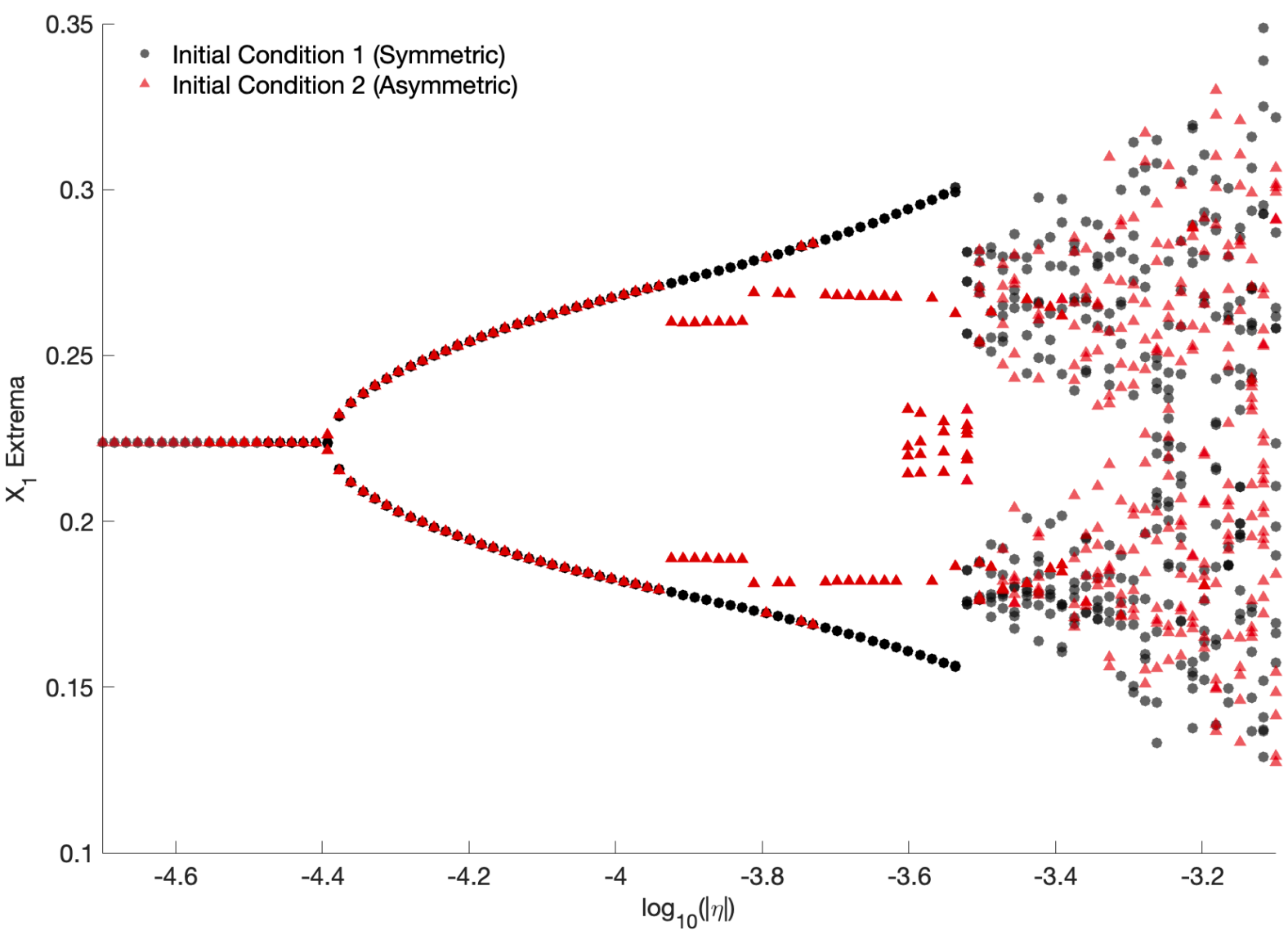

Figure 1: bifurcation diagram of the field amplitude extrema of laser 1($X_1$) in a complex coupled three laser arrays with π/11 coupling phase. The black and red dots represent sweeping with different initial condition.

## 3. Numerical simulation for Synchronized Chaos

We numerically integrated the coupled mode equations described in the previous section using a stiff ODE solver (ode15s in matlab). The system parameters in this paper were chosen to model typical semiconductor laser arrays, with a linewidth enhancement factor of α=5, pump p=0.05, detuning $\Delta = 0$, and a carrier-to-photon lifetime ratio of T=2000 unless otherwise specified. The initial conditions are chosen to be Z=0, X=0.01. The initial phases are random yet symmetric. The synchronized chaos states shown in this paper are confirmed through numerical simulation up to tens of microseconds, more than seven orders of magnitude longer than the photon lifetime.

### *3.1 Array with Fully Symmetrical Setup*

We start by examining a three-laser array with complex coupling. The time-domain simulation reveals that the outer lasers (Laser 1 and Laser 3) exhibit identical temporal behavior, while the center laser (Laser 2) follows a distinct chaotic trajectory. This is confirmed by the phase portrait of $X_1$ versus $X_3$, which collapses onto a diagonal line, indicating perfect synchronization (zero phase difference and identical amplitude) between the symmetrically located pair.

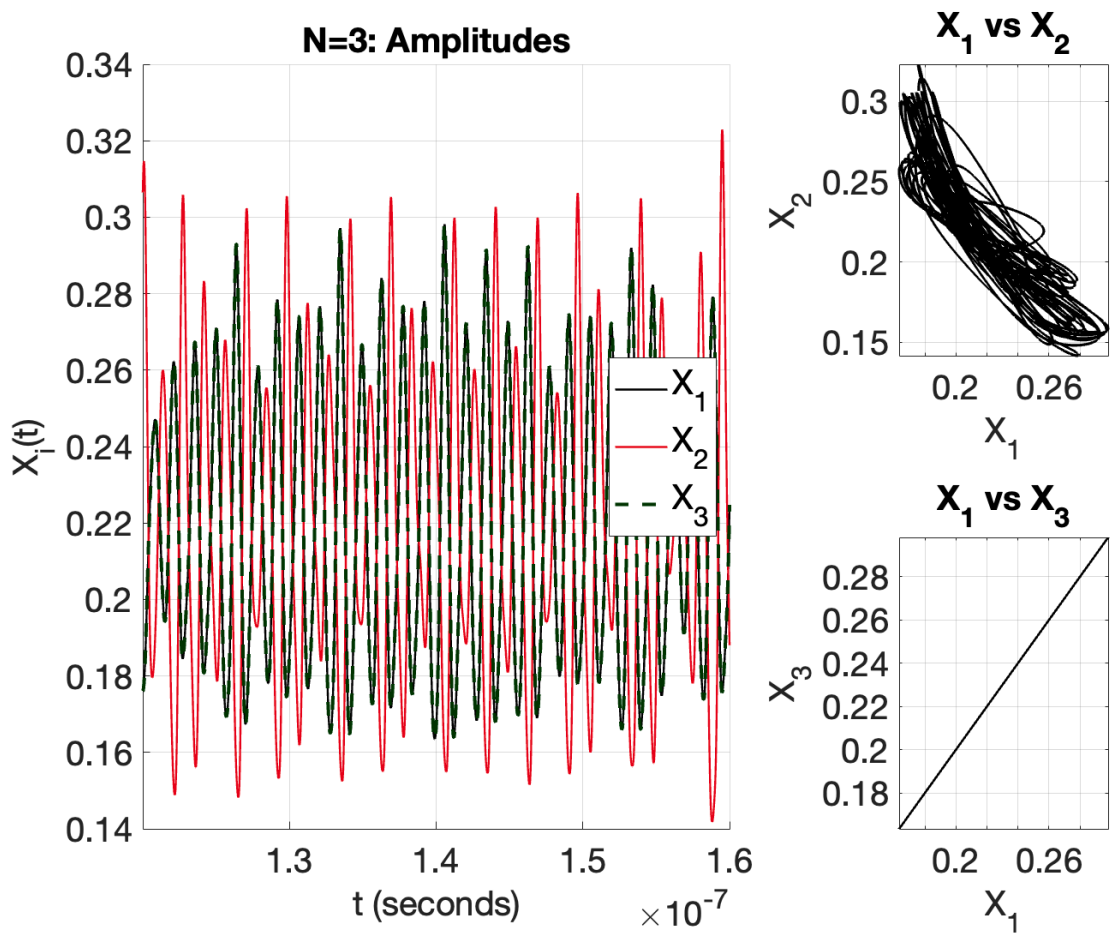

FIG 2: Time domain plot and phase diagram of three laser complex-coupled synchronized chaos, with $\eta = 10^{-3.4} e^{i\frac{\pi}{11}}$

Next we extend this simulation to larger arrays to test scalability. Simulations of 5, 7, and up to 11 lasers demonstrate that synchronized chaos persists in these higher-dimensional systems.

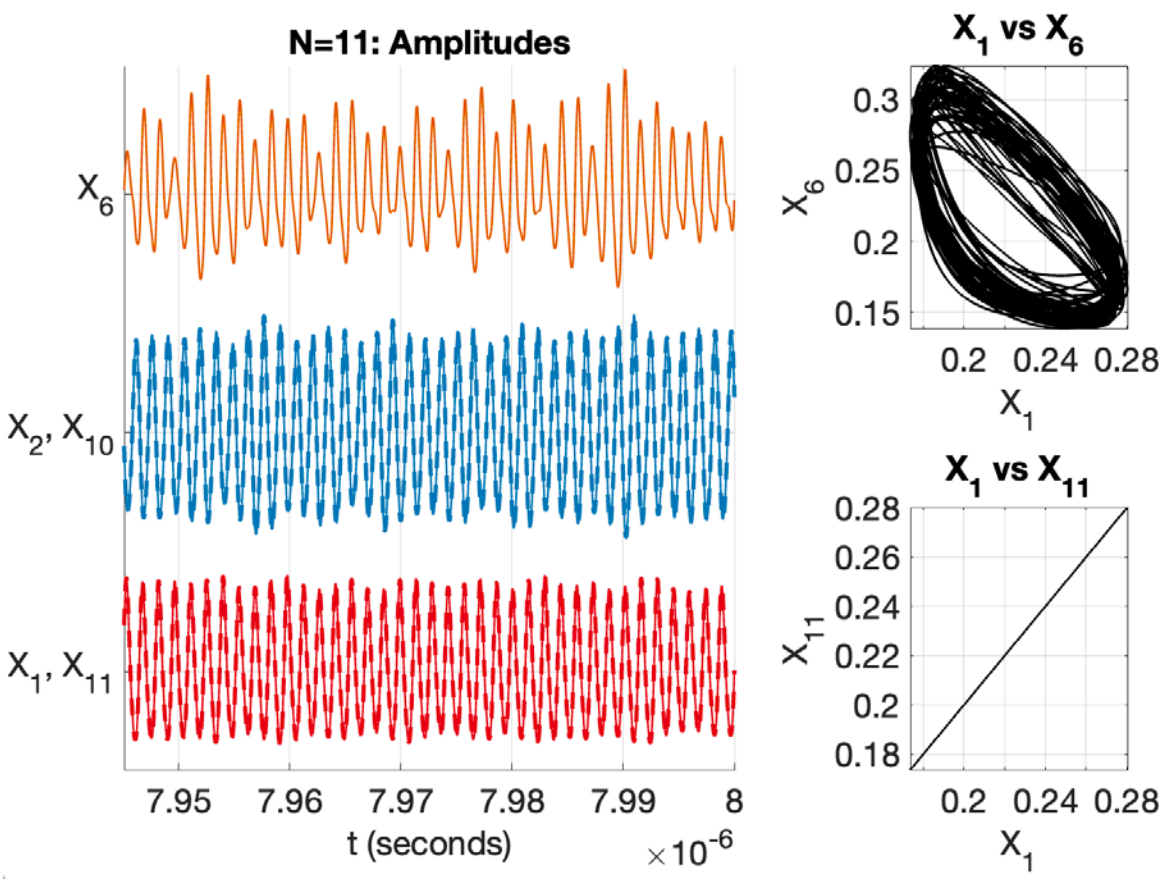

FIG 3: Time domain plots and phase diagrams showing a case of eleven coupled lasers with $\eta = 10^{-3.6} e^{i\frac{\pi}{11}}$ demonstrating a synchronized chaos state. Here $X_{11}$ and $X_{10}$ (dashed curves) perfectly synchronize with $X_1$ and $X_2$ (solid curve) respectively

However, synchronized chaos states in larger arrays exhibit a strong dependence on initial conditions. Our simulations indicate that achieving a synchronized chaos state relies heavily on the specific choice of initial phases. Furthermore, the synchronized chaos basin of attraction demonstrates a shrinking trend as the number of coupled lasers increases. Consequently, a randomly generated initial condition is significantly more likely to evolve into a synchronized chaos state for a 3-laser system than for an 11-laser array.

### *3.2 Array with small disorder*

To demonstrate the physical robustness of this phenomenon, we introduced built-in disorder in pumping strength and frequency detuning to simulate imperfections. We first investigated a three-laser complex coupled array($\psi = \frac{\pi}{11}$) subject to unequal pumping. The pump parameters were set to be spatially symmetrical and nonuniform. As shown in Fig. 4, in the presence of the built-in disorder, the array still showed stable synchronized chaos. We also introduced frequency detuning into a uniformly pumped three-laser complex-coupled array. The detuning parameters were varied on a scale that is 1-2 orders of magnitude smaller than the coupling strength(e.g.,$[\Delta_1, \Delta_2, \Delta_3] = [1,-1,1]$ * 1e-5). With these small deviations in natural frequencies, stable synchronized chaos survived.

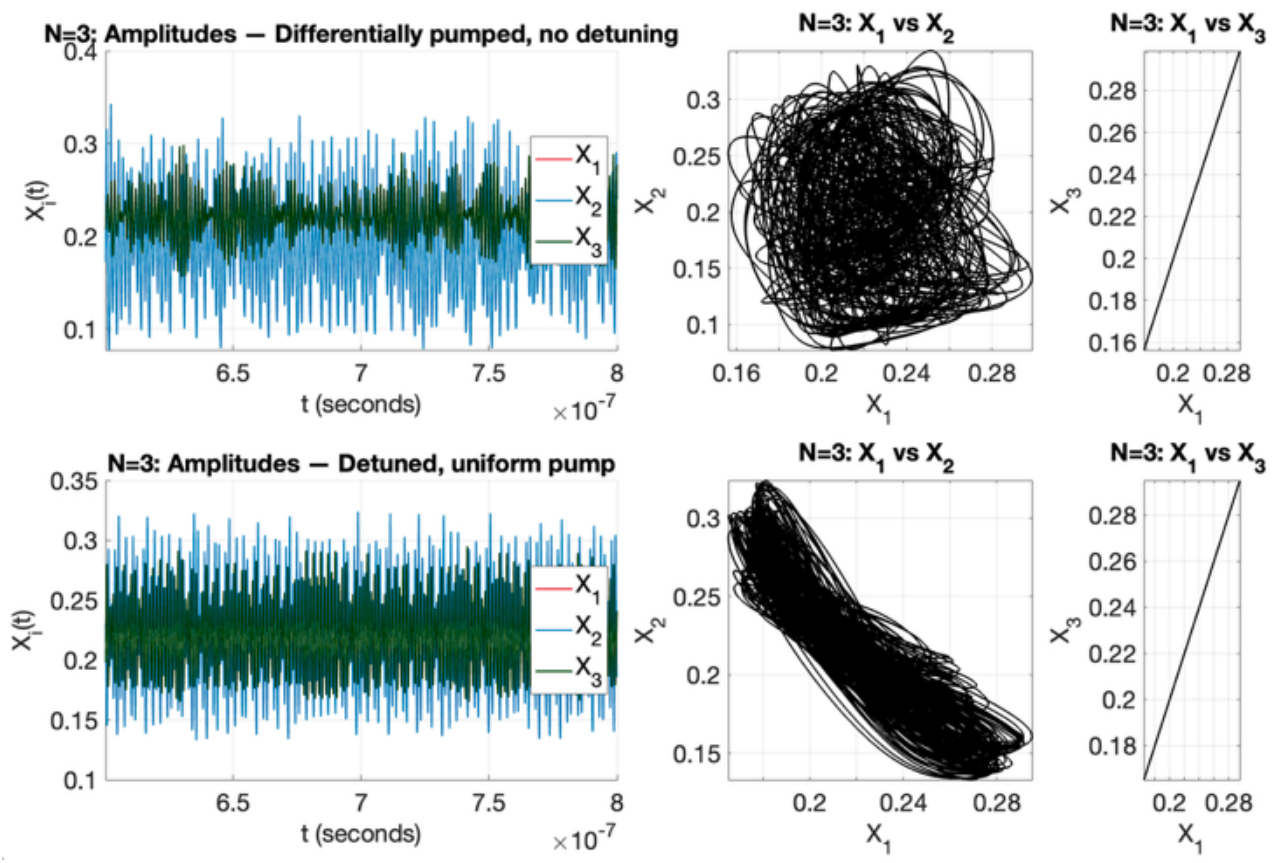

FIG 4: three-laser complex-coupled array with built-in disorder. Top row: Dynamics under a symmetrical nonuniform pump with p = [0.05 0.04 0.05], $\eta = 10^{-3.4} e^{i\pi/11}$ without detuning. Bottom row: Dynamics under a uniform pump with detuning $\Delta = [1 - 1\ 1] * 10^{-5}$ , $\eta = 10^{-3.45} e^{i\pi/11}$

### *3.3 Lyapunov Exponents*

Previous discussions of synchronized chaos with purely real coupling have used Lyapunov exponents as a useful metric for degree of chaos [13]. The Lyapunov exponents are the eigenvalues of the Jacobian of the variational equations averaged along an orbit of the system and hence provide a measure of the average exponential divergence of nearby orbits.

We calculated the Lyapunov spectrum(LS) using the Wolf algorithm, then used the Kaplan-Yorke formula to determine its Lyapunov dimension (D) [14].

$$D = \ j + \frac{\lambda_1 + \lambda_2 + \cdots + \lambda_j}{\lambda_{j+1}}, where\ LS = \{\lambda_1, \lambda_2, \ldots, \lambda_N\}, and\ j = \max\left\{m\colon \sum_{i=1}^{m} LE_i \geq 0\right\}$$

The Lyapunov dimension acts as a rough measure of the effective degrees of freedom that the dynamical system evolves in [14]. In the case of the three-laser, complex coupled (π/11) array, the Lyapunov spectrum is (0.70,0,0,-0.03,-0.32,-0.35,-0.36, -0.40,-1.43) bits/ns and from this we find that the Lyapunov dimension is 6.0. This value suggests that the system's evolution is confined to no more than six dimensions. This reduction is consistent with the system's symmetry: while the total phase space for three emitters is nine dimensional (3×3), the identical synchronization of the outer emitters (1 and 3) effectively removes three degrees of freedom, constraining the dynamics to a six-dimensional subspace (9−3=6). Similarly, we were able to find the Lyapunov spectrum in a five laser system($\eta = 10^{-3.46} e^{i\frac{\pi}{11}}$) to be (0.48,0.04,0.02,-0.07,-0.08,-0.08,-0.15,-0.24,-0.28,-0.33,-0.35,-0.4,-0.45,-0.57,-0.97) bits/ns. This gives a Lyapunov dimension of 7.8. This calculated dimension is in good agreement with our expectation since for 5 lasers with synchronization, 2 lasers will be redundant and act as an exact copy of the symmetrially located ones, resulting in an effective dimension of 3*5-3*2 = 9. The Lyapunov dimension for the 11-laser array is calculated to be 9.8. While this is well below the expected dimension of 18 (3×11−3×5), it is still a clear sign that the system is chaotic rather than quasi-periodic.

## 4. Transient Synchronized Chaos

While the preceding analysis demonstrated that a small coupling phase (π/11) preserves the robustness of synchronized chaos, increasing this coupling phase further could fundamentally change the system's dynamics. At larger coupling phase near $\pi/8$, we observe a transition to transient synchronized chaos. Although transient synchronized chaos has been reported previously in a three-element complex-coupled array [3], here we take a more detailed look into the statistical property of this phenomenon. In this regime, trajectories evolving from random initial conditions are sometimes attracted to a synchronized chaotic attractor. The system then remains perfectly synchronized for a sub-microsecond duration(on the order of $10^6$ times of photon-lifetime) before breaking spatial symmetry and decaying into an asynchronous state. This phenomenon shares core characteristics with classical transient chaos [11,15] but exhibits an additional layer of complexity due to spatial symmetry.

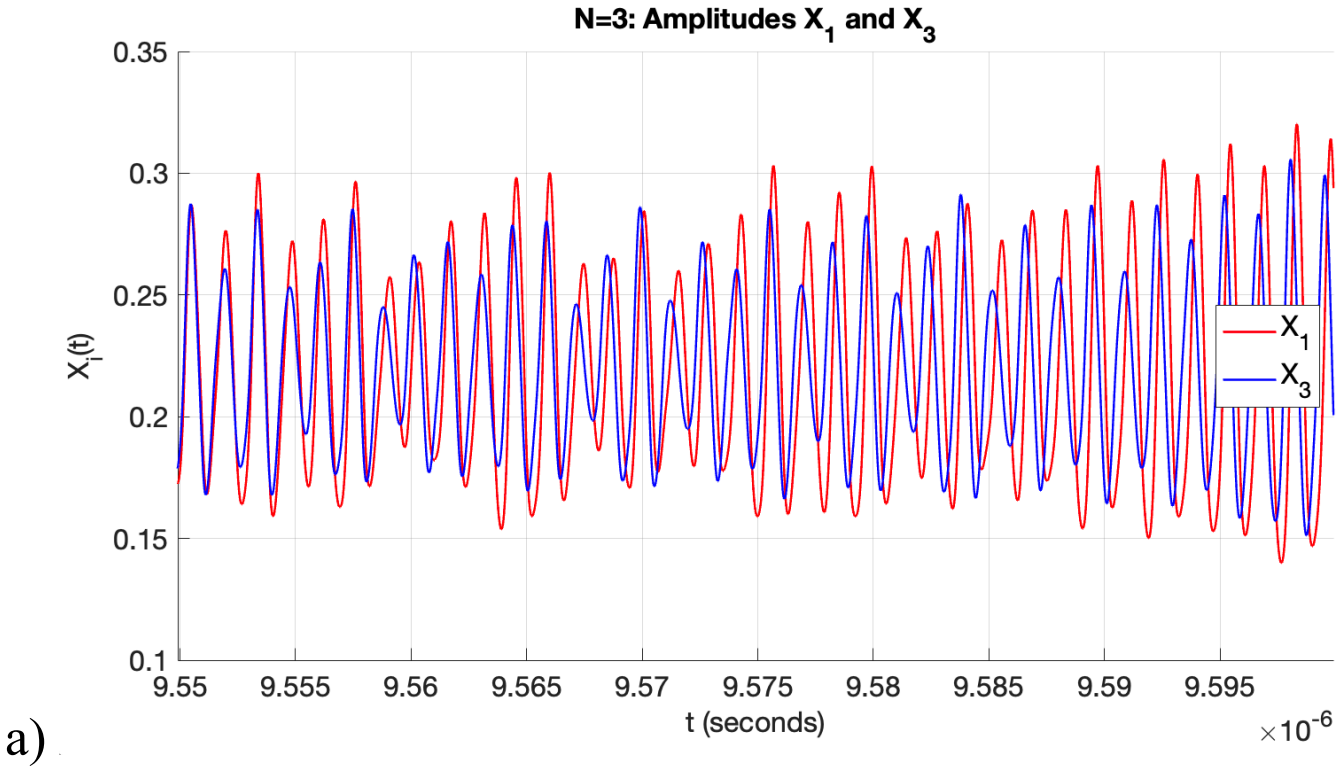

a)

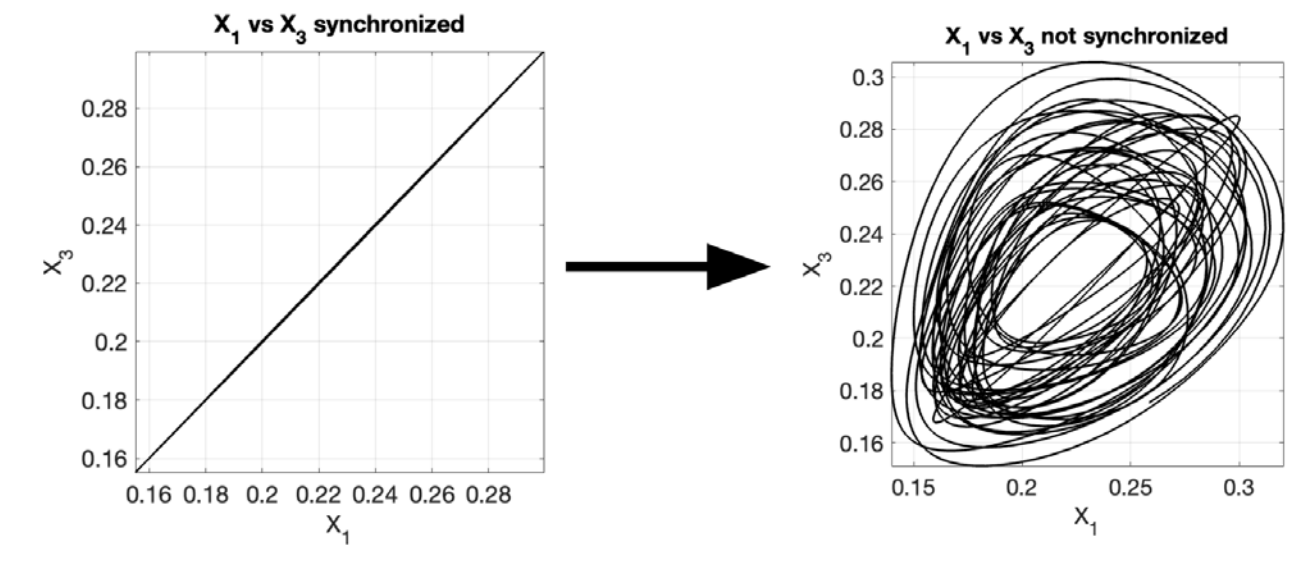

b)

FIG 5: three complex coupled array with coupling parameter $\eta = 10^{-3.4}e^{\frac{i\pi}{8}}$ The system initially exhibit synchronized chaos dynamics(X1=X3). After extended integration, system loses its spatial symetry(X1≠ X3) and evolves into a chaotic asynchronous state, we refer to this behavior as transient synchronized chaos a) time domain field amplitude b) phase portrait evolution

Figure 5 shows the time series of a 3-laser complex-coupled array with $\eta = 10^{-3.4}e^{\frac{i\pi}{8}}$ evolving from perfectly synchronized chaos to an asynchronous chaotic state. We refer to this phenomenon as transient synchronized chaos. To rigorously characterize this transient behavior and rule out numerical artifacts, we analyzed the statistical properties of a three-element array coupled at a coupling phase of π/8, adopting established methods for transient chaotic systems [11]. The system was initialized using an ensemble of finite-volume initical conditions near equilibrium. Specifically, the initial amplitudes $X_0$ were set to a common value $\sqrt{p}$ with an added independent random perturbation on the order of $10^{-3}$, i.e. $X_{0,i} = \sqrt{p} + 10^{-3}u_i, u_i \sim U(0,1)$. The carrier densities $Z_0$ were asymmetrically sampled from the interval $[0,\ 10^{-3}]$, i.e. $Z_{0,i} \sim U(0, 10^{-3})$. Finally, the phases $\phi$ were symmetrically distributed over [0,2π], subject to an asymmetric random perturbation bounded by π/10, i.e. $\phi_{0,k} = u_j + \frac{\pi}{10}u_k, ; u_j = u_{N+1-j} \sim U(0,2\pi)\ , u_k \sim U(0,1)$. With the exception of the coupling phase, all system parameters are identical to those of the three-laser case described earlier (Figure 2).

Because the initial conditions sampled here are asymmetrical, time is required before the system evolves into a fully synchronized state. We then identify the first time interval during which the array remains synchronized for longer than $10^2$ relaxation oscillation periods. Once this synchronized state is established, we record the first subsequent time at which the system departs from synchrony and becomes asynchronous, using the criterion $\phi_1 - \phi_3 > 10^{-7} rad$. The escape time is defined as the time between the end of this initial $10^2$ relaxation oscillation synchronized interval and the first subsequent desynchronization event, whose distribution is plotted on Figure 6a). We started from 640 randomly generated initial conditions using the procedure described above. Since our objective is to study the transient dynamics of the synchronized attractor rather than the probability of reaching it, the survival probability is normalized by the conditioned sample size, excluding trajectories that never enter the synchronized attractor within the integration window.

In standard transient chaos, the survival probability N(t) of trajectories within the chaotic region decays exponentially over time ( $N(t) \propto e^{-\kappa t}$ ) [11]. Our system exhibits a distinct bi-exponential decay as shown in Figure 6. We attribute this modified statistical distribution to the possible coexistence of two non-attracting chaotic sets with different lifetimes [11] in our complex-coupled three laser system.

To confirm the generality of this bi-exponential distribution, we evaluated the system at adjacent coupling phases from π/10 to π/8, all of which exhibit clear bi-exponential lifetime distributions shown in Figure 6 b). We also note that in the small coupling phase region studied, the decay rate is a monotonic increasing function of the coupling phase. This shows the tunability of the transient chaos escape time and could potentially act as a tunable platform for further transient synchronized chaos study.

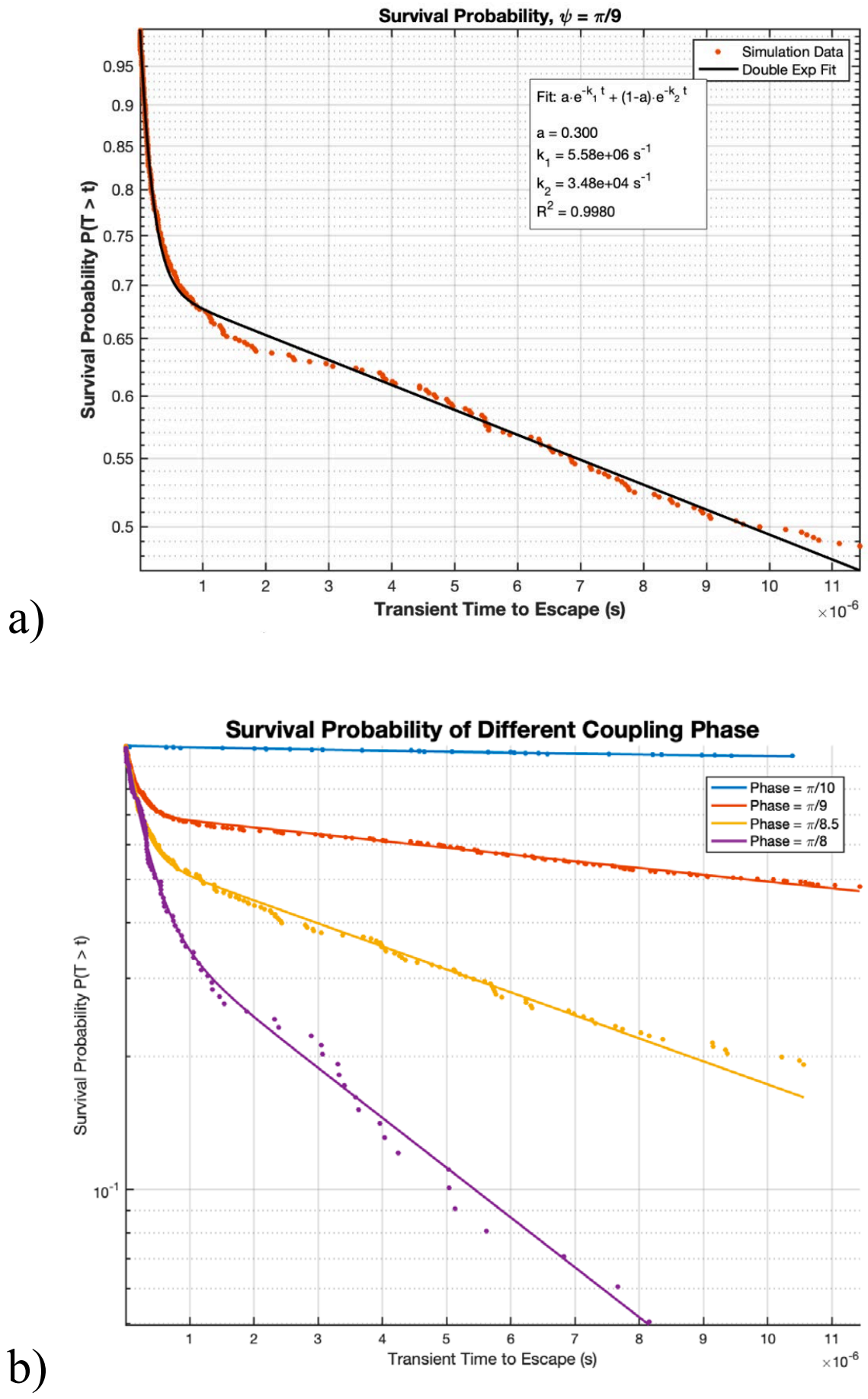

FIG 6: a) Survival probability N(t) for a coupling phase $\frac{\pi}{9}$ with fitted double exponential curve(solid) of the form $Ae^{-k_1 t} + (1 - A)e^{-k_2 t}$.b) Comparison of survival probability across different coupling phase with same coupling strength. The dots are transient time distributions obtained numerically, and the solid lines indicate the fitted double-exponential curves. A clear increase in both the survival probability and the decay rate is observed as the coupling phase increases.

## 5. Conclusion

In this work, we have performed a systematic numerical investigation into the scalability and robustness of synchronized chaos in coupled semiconductor laser arrays. While previous studies have largely focused on small systems (N=3), our results demonstrate that synchronized chaotic dynamics can persist in larger arrays of up to N=11 lasers.

By computing Lyapunov exponents and Lyapunov dimensions, we confirmed that these states are chaotic rather than quasi-periodic, while still maintaining perfect synchronization between symmetrically located laser pairs. Furthermore, we assessed the physical feasibility of these states by introducing realistic built-in disorder. Synchronized chaos is sustained even in the presence of  unequal pumping and small frequency detunings.

In addition, we identify a regime of transient synchronized chaos at larger coupling phases, in which trajectories remain on the synchronized chaotic state for a long but finite time before escaping to asynchronous states. The lifetime statistics are described by a bi-exponential survival distribution. The transient lifetime is also strongly tunable by the coupling phase. This combination of scalability, robustness to disorder, and controllable transient behavior establishes complex-coupled semiconductor laser arrays as a promising platform for synchronized chaos generation and for the exploration of high-dimensional transient nonlinear dynamics.

This material is based on work supported by Office of Naval Research under award number MURI ONR-N000142412548.